\documentclass[letterpaper]{article} 
\usepackage{aaai2026}  
\usepackage{times}  
\usepackage{helvet}  
\usepackage{courier}  
\usepackage[hyphens]{url}  
\usepackage{graphicx} 
\usepackage{natbib}  
\usepackage{caption} 
\usepackage{multirow}
\usepackage{tabularx} 
\usepackage{booktabs} 
\usepackage{adjustbox} 
\usepackage{amsmath} 
\usepackage{makecell} 
\usepackage{xurl} 
\usepackage[table]{xcolor}         
\usepackage{amsfonts}

\usepackage{algorithm}
\usepackage{algorithmic}

\usepackage{newfloat}
\usepackage{listings}
\DeclareCaptionStyle{ruled}{labelfont=normalfont,labelsep=colon,strut=off} 
\floatstyle{ruled}
\newfloat{listing}{tb}{lst}{}
\floatname{listing}{Listing}
\title{DINO-Mix: Distilling Foundational Knowledge with Cross-Domain CutMix for Semi-supervised Class-imbalanced Medical Image Segmentation}
\author{
    Xinyu Liu\textsuperscript{1}, Guolei Sun\textsuperscript{2}
}
\affiliations{
    \  
    \textsuperscript{\rm 1}The Chinese University of Hong Kong\\
    \textsuperscript{\rm 2}College of Computer Science, Nankai University\\
    {xinyuliu@link.cuhk.edu.hk}, {guolei.sun@nankai.edu.cn}
}

\usepackage{bibentry}

\begin{document}

\maketitle
\begin{abstract}
Semi-supervised learning (SSL) has emerged as a critical paradigm for medical image segmentation, mitigating the immense cost of dense annotations. However, prevailing SSL frameworks are fundamentally “inward-looking", recycling information and biases solely from within the target dataset. This design triggers a vicious cycle of confirmation bias under class imbalance, leading to the catastrophic failure to recognize minority classes. To dismantle this systemic issue, we propose a paradigm shift to a multi-level “outward-looking" framework. Our primary innovation is Foundational Knowledge Distillation (FKD), which looks outward beyond the confines of medical imaging by introducing a pre-trained visual foundation model, DINOv3, as an unbiased external semantic teacher. Instead of trusting the student’s biased high confidence, our method distills knowledge from DINOv3's robust understanding of high semantic uniqueness, providing a stable, cross-domain supervisory signal that anchors the learning of minority classes. To complement this core strategy, we further look outward within the data by proposing Progressive Imbalance-aware CutMix (PIC), which creates a dynamic curriculum that adaptively forces the model to focus on minority classes in both labeled and unlabeled subsets. This layered strategy forms our framework, DINO-Mix, which breaks the vicious cycle of bias and achieves remarkable performance on challenging semi-supervised class-imbalanced medical image segmentation benchmarks Synapse and AMOS.
\end{abstract}
\section{Introduction}
The accurate delineation of anatomical structures and pathologies in medical imaging is a cornerstone of modern diagnostics, treatment planning, and clinical research. While deep learning models have demonstrated remarkable capabilities for this task  \cite{li2025u}, their performance is heavily contingent on vast amounts of meticulously annotated data. The creation of such datasets is a notorious bottleneck, requiring expert knowledge and significant time. This reality has propelled semi-supervised learning (SSL) to the forefront of medical imaging research \cite{han2024deep}, offering a promising path to leverage large, readily available cohorts of unlabeled data alongside small, annotated sets.

Despite its promise, the dominant paradigm in semi-supervised learning is architecturally “inward-looking." Methodologies ranging from classic consistency-based approaches \cite{yu2019uamt, wu2022ssnet, luo2021urpc, most} to more recent, sophisticated frameworks all manage to refine information generated entirely from the target dataset's statistics. This design exposes a fatal vulnerability in the face of class imbalance, which is a condition endemic to medical imaging, where small lesions or organs are vastly outnumbered by background tissues or organs with large size. Although some recent works have been proposed to address the class-imbalance issue with selective self-training \cite{wei2021crest} or co-training \cite{wang2023dhc, zhang2025semantic}, they remain operating within the inward-looking paradigm, and would suffer from the vicious cycle of confirmation bias \cite{arazo2020pseudo}. This process begins with an initial model preference for the over-represented majority class. The model then generates pseudo-labels predicated on its own biased confidence, which are overwhelmingly skewed. Eventually, the model trained on its own prejudiced predictions will further cement its initial bias and progressively starve minority classes of supervisory signals until they are completely ignored.

We posit that any attempt to mitigate this cycle from within is fundamentally limited. The solution requires a paradigm shift: from an “inward-looking" to an “outward-looking" philosophy that introduces objective, external knowledge immune to the dataset's skewed statistics. Our primary contribution, Foundational Knowledge Distillation (FKD), which actualizes this philosophy by looking outward beyond the entire domain of medical imaging. We introduce a pre-trained visual foundation model DINOv3 \cite{simeoni2025dinov3} as a frozen unbiased external semantic teacher. Having learned from millions of diverse, natural images via self-supervision, DINO's representations are not tied to any specific downstream task or class distribution. 
Its understanding is based on universal principles of visual distinctiveness—texture, shape, and structure. By distilling its robust feature representations into our segmentation student, we provide a stable, objective supervisory signal. This external anchor, grounded in visual salience rather than class frequency, directly counteracts the vicious cycle. When the student model is uncertain about a rare but visually distinct lesion, our foundational model provides guidance with a strong, persistent gradient, ensuring the student learns its features.

To complement the powerful cross-domain guidance from the DINO teacher, we introduce a second, input-level “outward-looking” mechanism designed to build a more receptive student model named Progressive Imbalance-aware CutMix (PIC), which institutes a dynamic learning curriculum that evolves over the course of training. In the initial stages, when the model is most susceptible to the dataset's inherent bias, PIC operates in a heavily class-aware mode. It aggressively samples and pastes regions from minority classes, forcing the student to immediately confront and prioritize learning from the most challenging and underrepresented examples. As training matures and the model develops a more balanced understanding, PIC progressively relaxes this constraint, transitioning its sampling strategy towards a uniform distribution. This strategic shift prevents overfitting on minority classes and ensures the student builds robust features for the entire dataset. By first tackling the core data imbalance and then generalizing, PIC cultivates a more competent student, ideally prepared to absorb and leverage the profound knowledge distilled from its foundational teacher.

The synergy between these two components forms a cohesive and powerful framework named \textit{DINO-Mix}. The external teacher provides the unbiased \textit{what} to learn, breaking the semantic bias loop, while the adaptive intra-domain regularizer creates a more robust student that is better at \textit{how} to learn, especially for rare classes. This two-pronged “outward-looking" attack dismantles the vicious cycle of confirmation bias at its root. Through extensive experiments on the highly imbalanced Synapse and AMOS datasets, we demonstrate that our approach not only resolves a critical failure mode of existing SSL methods but also sets a new state-of-the-art in semi-supervised medical image segmentation. The main contributions of our work include:
\begin{itemize}
    \item We propose a novel “outward-looking" paradigm Foundational Knowledge Distillation (FKD) for semi-supervised medical segmentation by leveraging a pre-trained foundation model DINOv3 as an external, unbiased semantic teacher, which provides a stable supervisory signal for minority classes.
    \item We introduce Progressive Imbalance-aware CutMix (PIC), a dynamic intra-domain regularization strategy that constructs a learning curriculum, beginning with a class-aware focus and progressively shifting to uniform sampling to build a robust and generalized student model.
\item We combine these components into a cohesive framework, DINO-Mix, which achieves new state-of-the-art results on challenging and highly imbalanced semi-supervised segmentation benchmarks.
\end{itemize}

\section{Related Work}

\subsection{Semi-supervised Medical Image Segmentation}

Semi-supervised medical image segmentation (SSMIS) addresses the scarcity of labeled medical data, especially in 3D imaging like ultrasound and MRI, by leveraging large pools of unlabeled scans \cite{yu2019uamt, luo2021dtc, luo2021urpc, most, liu2024diffrect, chen2021cps}. The field has evolved along two main, often intertwined, approaches: consistency regularization and pseudo-labeling. In consistency regularization, models are trained to produce invariant predictions on perturbed inputs of the same unlabeled data, often within a teacher-student framework. Methods like Mean Teacher \citep{tarvainen2017meanteacher} and its medical adaptation UA-MT \citep{yu2019uamt} employ EMA and uncertainty weighting, while URPC \citep{luo2021urpc} and DTC \citep{luo2021dtc} introduce multi-level feature and dual-task constraints to improve boundary quality. The pseudo-labeling branch generates targets from the model's own confident predictions, as seen in FixMatch \cite{sohn2020fixmatch}. To mitigate error accumulation, co-training architectures like Cross Pseudo Supervision (CPS) \cite{chen2021cps} were proposed, where two networks provide mutual supervision to correct each other's mistakes. Some recent approaches have been exploring more diverse data transformations including copy-paste augmentation \cite{bai2023bcp} and frequency domain interpolation \cite{hu2025beta}. However, these methods are fundamentally "inward-looking", as they rely entirely on supervisory signals generated from within the training process itself. This could create a vulnerability to confirmation bias when extended to multi-organ segmentation tasks, where an initial preference for majority classes is rapidly amplified, leading to the neglect of minority structures.

\begin{figure*}
    \centering
    \includegraphics[width=0.99\linewidth]{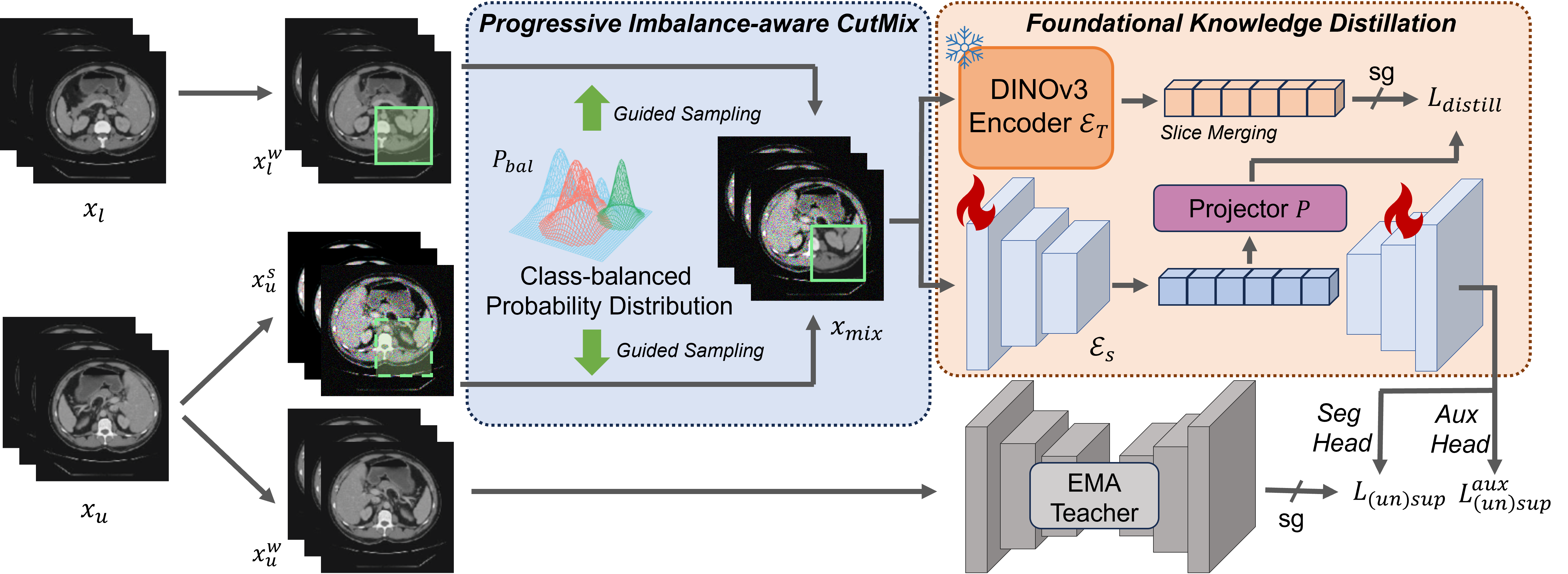}
    \caption{The overview of our proposed DINO-Mix framework for semi-supervised class-imbalanced medical image segmentation during training, which includes two main components Progressive Imbalance-aware CutMix (PIC) and Foundational Knowledge Distillation (FKD).}
    \label{fig:main}
\end{figure*}

\subsection{Semi-supervised Class-imbalanced Learning}

A fundamental challenge in real-world medical imaging is class imbalance, which impairs model generalization, especially for rare classes \cite{wang2023dhc, li2024semantic, zhang2025semantic, sohn2020fixmatch, yang2022semi}. This issue is magnified in semi-supervised learning, which is due to the scarcity of labeled examples for minority classes. Recent research has developed specialized SSL methods to tackle this challenge. Some focus on creating a balanced training set using pseudo-labels, such as CReST \cite{wei2021crest}, which explicitly selects pseudo-labeled data to favor minority classes. Others adjust the learning dynamics directly: CLD \cite{lin2022cld} re-weights the loss based on class instance counts, while DHC \cite{wang2023dhc} advances this by considering both class distribution and learning difficulty. GenericSSL \cite{wang2023towards} extracts distribution-invariant features from aggregated information from multiple distributions/domains. SKCDF \cite{zhang2025semantic} adopts labeled data to guide the generation of pseudo labels in feature level via a cross-attention mechanism. However, these approaches are confined to the data distributions within the dataset itself, which may be still restricted by the limited semantics and representations.

\subsection{Leveraging Foundational Models for Semi-supervised Learning}

The advent of foundational models, pre-trained on vast labeled or unlabeled datasets, has revolutionized the field of computer vision by learning robust and highly generalizable representations \cite{clip, oquab2023dinov2, simeoni2025dinov3}. Their profound ability to understand semantic content from unlabeled data makes them exceptionally well-suited for SSL, where the scarcity of labeled data is a primary challenge  \cite{li2024vclipseg, huang2025text, pham2025semi, hoyer2024semivl, li2025clip}. 
This methodology has been validated across diverse and complex domains. For instance, Hoyer \textit{et al.} \cite{hoyer2024semivl} proposed SemiVL, a method that integrates rich semantic priors from a CLIP model to resolve ambiguities between visually similar classes, a common failure point for traditional SSL that rely solely on visual consistency. The application of this principle extends even to more nuanced scenarios, such as fine-grained open-set SSL, where Li \textit{et al.} \cite{li2025clip} employ a CLIP-driven, coarse-to-fine framework to progressively focus on distinctive visual details. In the task of 3D semantic scene completion, Pham \textit{et al.} \cite{pham2025semi} introduce a framework that utilizes depth and segmentation foundation models to extract geometric and semantic 3D clues from unlabeled images. 
However, a common thread in these pioneering efforts is their significant reliance on vision-language models (e.g., CLIP) or task-specific foundation models. While highly effective, this focus has left the potential of purely vision-based self-supervised foundational models, which learn powerful semantic representations through self-supervised objectives without any textual data or task-specific labels. They are relatively underexplored in the context of SSL guidance. Models like DINO family, for instance, have shown an exceptional ability to learn object-level features and semantic segmentation maps directly from images. Yet, their direct integration as semantic teachers within semi-supervised frameworks remains a nascent area.

\section{Methodology}
Our DINO-Mix framework is built upon a consistency regularization paradigm \cite{sohn2020fixmatch, most} with an Exponential Moving Average (EMA) teacher model to generate stable pseudo-labels, as illustrated in Fig. \ref{fig:main}. For unlabeled data, we generate a weakly-augmented view and a strongly-augmented view. The EMA teacher model produces a high-confidence pseudo-label from the weakly-augmented view, which then serves as a supervision target for the prediction from the strongly-augmented view, enforcing model consistency. Moreover, an auxiliary balancing classifier \cite{lee2021abc, zhang2025semantic} is applied to disentangle the learning of general representations from the specific task of handling class imbalance. Into this baseline, we integrate a two-pronged "outward-looking" strategy. First, we introduce an external, pre-trained foundation model as a source of unbiased semantic knowledge to break the cycle of confirmation bias. Second, we employ a novel intra-domain data mixing strategy to regularize the student model and enhance its generalizability.

\subsection{Foundation Knowledge Distillation (FKD)}
The fundamental flaw of conventional semi-supervised learning is its susceptibility to a vicious cycle of confirmation bias. When faced with class imbalance, a student model's initial preference for the majority class leads it to generate biased pseudo-labels. Training on these self-generated flawed targets causes the model to become increasingly confident in its own errors, progressively starving minority classes of a supervisory signal until they are ignored entirely. To break this destructive feedback loop, an "outward-looking" approach that introduces a source of supervision completely external to the student model and immune to the dataset's skewed statistics.
To realize this external semantic anchor, we employ a pre-trained DINOv3 Vision Transformer, with its encoder denoted as $\mathcal{E}_T$, which remains frozen throughout training. Having learned from millions of diverse, non-medical images, the pretrained DINOv3 encoder $\mathcal{E}_T$ possesses a rich, task-agnostic understanding of fundamental visual concepts, making it an ideal objective teacher. We transfer this knowledge to our 3D segmentation student, whose encoder is denoted as $\mathcal{E}_S$, via feature-level distillation on an unlabeled 3D volume $\boldsymbol{x_u}$. First, we pass the data through the student encoder to extract its deepest feature map, $\boldsymbol{F_S} = \mathcal{E}_S(\boldsymbol{x_u}) \in \mathbb{R}^{C_S \times D_S \times H_S \times W_S}$. To obtain a corresponding representation from the 2D teacher, we process the volume slice-by-slice, stacking the resulting deep feature maps from $\mathcal{E}_T$ into a 3D teacher volume, $\boldsymbol{F_T}$. To harmonize these representations, we spatially align $\boldsymbol{F_T}$ via adaptive average pooling and use a lightweight 3D convolutional projector, $\mathcal{P}$, to map the student's feature channels $C_S$ to the teacher's dimension $C_T$. The distillation is driven by minimizing the mean squared error between their L2-normalized features, defining our distillation loss $\mathcal{L}_{distill}$ as:
\begin{equation}
\mathcal{L}_{distill} = \left| \frac{\mathcal{P}(\mathcal{E}_S(\boldsymbol{x_u}))}{|\mathcal{P}(\mathcal{E}_S(\boldsymbol{x_u}))|_2} - \frac{\mathrm{sg}(\boldsymbol{F_T})}{|\mathrm{sg}(\boldsymbol{F_T})|_2} \right|_2^2
\label{eq:distill}
\end{equation}
where $\mathrm{sg}(\cdot)$ denotes the stop-gradient operator, ensuring the teacher's weights remain frozen.

The mechanism of this distillation loss directly subverts the confirmation bias cycle. Consider a rare but visually distinct structure where the student's predictive confidence is low. While a conventional SSL method would generate a weak or incorrect pseudo-label, our DINOv3 teacher, recognizing the region's unique texture and shape, provides a strong and consistent feature representation. Minimizing $\mathcal{L}_{distill}$ therefore creates a powerful and persistent gradient that forces the student to learn meaningful features for this minority class. This external semantic guidance ensures that all visually distinct regions receive robust supervision, fundamentally preventing the catastrophic representational collapse that plagues inward-looking SSL frameworks.

\subsection{Progressive Imbalance-aware CutMix (PIC)}
A significant challenge in multi-organ medical image segmentation is class imbalance, where standard data augmentations fail to provide sufficient exposure to rare classes. To address this, we propose Progressive Imbalance-aware CutMix (PIC), a dynamic data mixing strategy that adapts its behavior as the model trains. The core principle is to primarily focus on the under-represented classes, forcing the model to refine its decision boundaries for these critical, tail classes. When the model's capability of recognizing all classes is well established, the exploration is gradually shifted to class-agnostic. Therefore, PIC samples patches uniformly from all classes, allowing the model to learn stable, general features. 
This progressive shift is governed by a carefully formulated sampling distribution. First, we quantify the class imbalance from the labeled dataset. Let $N_c$ be the total pixel count for class $c$ in a representative subset of the labeled data. We compute an Imbalance Ratio (IR) for each class as:
\begin{equation}
I_c = \frac{\min_{j} N_j}{N_c}.
\end{equation}
This assigns a value of $1.0$ to the rarest class and smaller values to more frequent ones. From this, we derive a static, class-balanced probability distribution $P_{bal}$ that emphasizes tail classes, controlled by a focusing parameter $\gamma$:
\begin{equation}
\label{eq:p_bal}
P_{bal}(c) = \frac{(I_c)^\gamma}{\sum_{j=0}^{C-1} (I_j)^\gamma},
\end{equation}
where $C$ is the total number of classes. A higher $\gamma$ further sharpens the distribution towards minority classes.
The PIC module dynamically interpolates between $P_{bal}$ and a uniform distribution $P_{uni}$ as training evolves. The final sampling probability distribution $P_E$ at epoch $E$ is defined as:
\begin{equation}
\label{eq:progressive_mix_final}
P_E = (1 - \alpha_t) \cdot P_{bal} + \alpha_t \cdot P_{uni},\  \alpha_t = \min\left(1, \frac{E}{\eta * E_{max}}\right).
\end{equation}
At the start of training ($E \approx 0$), the progressive factor $\alpha_t$ is near zero, making $P_E$ is governed by the class-balancing distribution $P_{bal}$. As $E$ approaches the scaled maximum number of epochs $\eta * E_{max}$, $\alpha_t$ converges to one, and the sampling follows the uniform distribution $P_{uni}$. The scaling hyperparameter $\eta$ controls the shifting speed, where a larger number leads to slower shifting towards the uniform distribution.
For each application, a voxel of a target class is sampled from $P_E$, and a rectangular patch region is formed with the sampled voxel as center. Then, we cut the region from a labeled image and pasted onto a strongly-augmented unlabeled image. This progressive shift creates a stable and effective learning trajectory, improving performance in class-imbalanced, semi-supervised environments.

\begin{table*}[t]
\caption{Results on Synapse dataset with 20\% labeled data. `General' or `Imbalance' indicates whether the methods consider the imbalance issue or not. Sp: spleen, RK: right kidney, LK: left kidney, Ga: gallbladder, Es: esophagus, Li: liver, St: stomach, Ao: aorta, IVC: inferior vena cava, PSV: portal \& splenic veins, Pa: pancreas, RAG: right adrenal gland, LAG: left adrenal gland. Results of 3-times repeated experiments are reported in the ’mean±std’ format. Best results are boldfaced.}
  \footnotesize
  \setlength{\tabcolsep}{1.3mm}
\label{sota_IBSSL}
\resizebox*{\linewidth}{!}{
\begin{tabular}{c|c|cc|ccccccccccccc}
\toprule
\multicolumn{2}{c|}{\multirow{2}{*}{Methods}}  & {Avg.} & {Avg.} & \multicolumn{13}{c}{Dice of Each Class}                                        \\ 
\multicolumn{2}{c|}{}                          &   Dice                         &   ASD                        & Sp   & RK   & LK   & Ga   & Es   & Li   & St   & Ao   & IVC  & PSV  & PA   & RAG  & LAG  \\\midrule

\multirow{9}{*}{\rotatebox{90}{General}}         & V-Net (fully)      & 62.09±1.2	&10.28±3.9	& 84.6	& 77.2	& 73.8	& 73.3	& 38.2	& 94.6	& 68.4	& 72.1	& 71.2	& 58.2	& 48.5	& 17.9	& 29.0 \\ \midrule

& UA-MT (MICCAI'19)   & 20.26±2.2	&71.67±7.4	& 48.2	& 31.7	& 22.2	& 0.0	& 0.0	& 81.2	& 29.1	& 23.3	& 27.5	& 0.0	& 0.0	& 0.0	& 0.0  \\
                 
& URPC (MICCAI'21)      & 25.68±5.1	&72.74±15.5	& 66.7	& 38.2	& 56.8	& 0.0	& 0.0	& 85.3	& 33.9	& 33.1	& 14.8	& 0.0	& 5.1	& 0.0	& 0.0  \\

& CPS (CVPR'21)     & 33.55±3.7	&41.21±9.1	& 62.8	& 55.2	& 45.4	& 35.9	& 0.0	&{91.1}	& 31.3	& 41.9	& 49.2	& 8.8	& 14.5	& 0.0	& 0.0  \\

& SS-Net (MICCAI'22)  & 35.08±2.8	&50.81±6.5	& 62.7	& 67.9	& 60.9	& 34.3	& 0.0	& 89.9	& 20.9	& 61.7	& 44.8	& 0.0	& 8.7	& 4.2	& 0.0  \\

& DST (NeurIPS'22)      & 34.47±1.6	&37.69±2.9	& 57.7	& 57.2	& 46.4	& 43.7	& 0.0	& 89.0	& 33.9	& 43.3	& 46.9	& 9.0	& {21.0}	& 0.0	& 0.0  \\

& DePL (CVPR'22)       & 36.27±0.9	&36.02±0.8	& 62.8	& 61.0	& 48.2	& 54.8	& 0.0	& {90.2}	& {36.0}	& 42.5	& 48.2	& 10.7	& 17.0	& 0.0	& 0.0  \\ \midrule

\multirow{9}{*}{\rotatebox{90}{Imbalance}} 
& Adsh (ICML'22)     & 35.29±0.5	&39.61±4.6	& 55.1	& 59.6	& 45.8	& 52.2	& 0.0	& 89.4	& 32.8	& 47.6	& 53.0	& 8.9	& 14.4	& 0.0	& 0.0  \\ 

& CReST (CVPR'21)      & 38.33±3.4	&{22.85±9.0}	& 62.1	& 64.7	& 53.8	& 43.8	& {8.1}	& 85.9	& 27.2	& 54.4	& 47.7	& 14.4	& 13.0	& {18.7}	& 4.6  \\

& SimiS (arXiv'22)      & 40.07±0.6	&32.98±0.5	& 62.3	& {69.4}	& 50.7	& 61.4	& 0.0	& 87.0	& 33.0	& 59.0	& {57.2}	& {29.2}	& 11.8	& 0.0	& 0.0  \\

& Basak (MICCAI'22)       &33.24±0.6	&43.78±2.5	& 57.4	& 53.8	& 48.5	& 46.9	& 0.0	& 87.8	& 28.7	& 42.3	& 45.4	& 6.3	& 15.0	& 0.0	& 0.0   \\
                                 
& CLD (MICCAI'22)      &{41.07±1.2}	&32.15±3.3	& 62.0	& 66.0	& {59.3}	& {61.5}	& {0.0}	& 89.0	& 31.7	& {62.8}	& 49.4	& 28.6	& 18.5	& {0.0}	& {5.0}  \\
 
& DHC (MICCAI'23)    & 48.61±0.9	&10.71±2.6	& 62.8	& 69.5	& 59.2	& {66.0}	& 13.2	& 85.2	& 36.9	& 67.9	& 61.5	& 37.0	& 30.9	& 31.4	& 10.6 \\ 

 &  {A\&D} (NeurIPS'23)    & {60.88±0.7}	&{2.52±0.4}	& {85.2} & {66.9} & {67.0}  & 52.7  & {62.9} &89.6 &{52.1} &{83.0} &{74.9} &{41.8} &{43.4} &{44.8} &{27.2} \\ 

 &  {SKCDF} (CVPR'25)   & {64.27±1.4}	&{1.45±0.1}	& 79.5 &72.1 &67.6 &59.8 &60.7 &93.3 &\textbf{61.7} &85.4 &78.5 &41.8 &\textbf{50.9} &46.4 &\textbf{37.8} \\ 
\rowcolor{cyan!20} & {DINO-Mix (Ours)}  & \textbf{{66.45±0.1}}	&\textbf{{1.27±0.1}}	&\textbf{89.0}& \textbf{73.3}& \textbf{71.8}& \textbf{68.7}& \textbf{64.0}& \textbf{94.4}& 59.4& \textbf{86.2}& \textbf{79.0}& \textbf{48.6}& 45.6& \textbf{47.3}& 36.6 \\
\bottomrule
\end{tabular}

}

\end{table*}

\subsection{Optimization}

The DINO-Mix framework is trained end-to-end by minimizing a composite loss function that integrates supervised signals, consistency regularization, and external knowledge distillation. The total loss $\mathcal{L}_{total}$ is defined as a weighted sum of three components:
\begin{equation}
    \mathcal{L}_{total} = \mathcal{L}_{sup} + \mathcal{L}_{sup}^{aux} + \mathcal{L}_{distill} + \lambda_{unsup} (\mathcal{L}_{unsup} + \mathcal{L}_{unsup}^{aux}) ,
\end{equation}
where $\lambda_{sup}$ is the hyperparameter that balances the unsupervised loss terms. The supervised loss, $\mathcal{L}_{sup}$ and $\mathcal{L}_{unsup}$ are calculated on labeled data $\boldsymbol{x_l}$ and the PIC mixed image $\boldsymbol{x}_{mix}$, with the segmentation loss (combined Dice and Cross-Entropy). $\mathcal{L}_{sup}^{aux}$ and $\mathcal{L}_{unsup}^{aux}$ are the masked segmentation loss that aims to balance the classes following \cite{zhang2025semantic}. Finally, the distillation loss $\mathcal{L}_{distill}$ is defined in Equation \ref{eq:distill}, which aligns the student's feature representation of $\boldsymbol{x}_{mix}$ with the frozen DINOv3 teacher's representation, thereby injecting unbiased semantic guidance into the training process of the model. 

\section{Experiments}

\subsection{Datasets}
\noindent \textbf{Synapse dataset}. The Synapse dataset provides 30 axial contrast-enhanced abdominal CT scans for multi-organ segmentation. It features 13 distinct foreground classes: spleen (Sp), right kidney (RK), left kidney (LK), gallbladder (Ga), esophagus (Es), liver (Li), stomach (St), aorta (Ao), inferior vena cava (IVC), portal \& splenic veins (PSV), pancreas (Pa), right adrenal gland (RAG), and left adrenal gland (LAG), alongside a ubiquitous background class. For consistency with prior work \cite{wang2023dhc, zhang2025semantic}, all scans were resampled to a uniform dimension of 80 × 160 × 160 voxels. The dataset was subsequently divided randomly into a training set of 20 scans, a validation set of 4 scans, and a test set of 6 scans. To ensure robustness against data sampling variability inherent to the limited dataset size, experiments on Synapse were conducted three times with independent random seeds.

\noindent \textbf{AMOS dataset}. The AMOS dataset presents a broader scope with 360 scans and a modified set of anatomical structures compared to Synapse. Specifically, it excludes the portal \& splenic veins (PSV) but incorporates duodenum (Du), bladder (Bl), and prostate/uterus (P/U) as new classes. The dataset's partitioning yields 216 scans for training, 24 for validation, and 120 for testing.

\begin{table*}[!t]
  \centering
  \footnotesize
  \caption{Results on AMOS dataset with 5\% labeled data. Sp: spleen, RK: right kidney, LK: left kidney, Ga: gallbladder, Es: esophagus, Li: liver, St: stomach, Ao: aorta, IVC: inferior vena cava, Pa: pancreas, RAG: right adrenal gland, LAG: left adrenal gland, Du: duodenum, Bl: bladder, P/U: prostate/uterus. Best results are boldfaced.}
  \setlength{\tabcolsep}{1.3mm}
   \resizebox*{\linewidth}{!}{
   \begin{tabular}{@{}cc|cc|cccccccccccccccc@{}}
    \toprule
    & \multirow{2}{*}{Method} & Avg. & Avg. & \multicolumn{13}{c}{Dice of Each Class}  \\
    & & Dice & ASD & Sp & RK & LK & Ga & Es & Li & St & Ao & IVC & Pa & RAG & LAG & Du & Bl & P/U \\
    \midrule
    \multicolumn{1}{c|}{} & VNet (fully) 2016 & 76.50 & 2.01 & 92.2 & 92.2 & 93.3 & 65.5 & 70.3 & 95.3 & 82.4 & 91.4 & 85.0 & 74.9 & 58.6 & 58.1 & 65.6 & 64.4 & 58.3 \\
    \midrule
    \multicolumn{1}{c|}{\multirow{6}{*}{\rotatebox[origin=c]{90}{General}}}  & UA-MT (MICCAI'19)  & 42.16  & 15.48  & 59.8 & 64.9 & 64.0 & 35.3 & 34.1 & 77.7 & 37.8 & 61.0 & 46.0 & 33.3 & 26.9 & 12.3 & 18.1 & 29.7 & 31.6 \\
    \multicolumn{1}{c|}{} & URPC (MICCAI'21) & 44.93 & 27.44 & 67.0 & 64.2 & 67.2 & 36.1 & 0.0 & 83.1 & 45.5 & 67.4 & 54.4 & 46.7 & 0.0 & 29.4 & 35.2 & 44.5 & 33.2 \\
    \multicolumn{1}{c|}{} & CPS (CVPR'21) & 41.08 & 20.37 & 56.1 & 60.3 & 59.4 & 33.3 & 25.4 & 73.8 & 32.4 & 65.7 & 52.1 & 31.1 & 25.5 & 6.2 & 18.4 & 40.7 & 35.8 \\
    \multicolumn{1}{c|}{} & SS-Net (MICCAI'22) & 33.88 & 54.72 & 65.4 & 68.3 & 69.9 & 37.8 & 0.0 & 75.1 & 33.2 & 68.0 & 56.6 & 33.5 & 0.0 & 0.0 & 0.0 & 0.2 & 0.2 & \\
    \multicolumn{1}{c|}{} & DST (NeurIPS'22) & 41.44 & 21.12 & 58.9 & 63.3 & 63.8 & 37.7 & 29.6 & 74.6 & 36.1 & 66.1 & 49.9 & 32.8 & 13.5 & 5.5 & 17.6 & 39.1 & 33.1 \\
    \multicolumn{1}{c|}{} & DePL (CVPR'22) & 41.97 & 20.42 & 55.7 & 62.4 & 57.7 & 36.6 & 31.3 & 68.4 & 33.9 & 65.6 & 51.9 & 30.2 & 23.3 & 10.2 & 20.9 & 43.9 & 37.7 \\
    \midrule
    \multicolumn{1}{c|}{\multirow{8}{*}{\rotatebox[origin=c]{90}{Imbalance}}}  & Adsh (ICML'22) & 40.33 
 & 24.53 & 56.0 & 63.6 & 57.3 & 34.7 & 25.7 & 73.9 & 30.7 & 65.7 & 51.9 & 27.1 & 20.2 & 0.0 & 18.6 & 43.5 & 35.9 \\
    \multicolumn{1}{c|}{} & CReST (CVPR'21) & 46.55 & 14.62 & 66.5 & 64.2 & 65.4 & 36.0 & 32.2 & 77.8 & 43.6 & 68.5 & 52.9 & 40.3 & 24.7 & 19.5 & 26.5 & 43.9 & 36.4 \\
    \multicolumn{1}{c|}{} & SimiS (arXiv'22) & 47.27 & 11.51 & 77.4 & 72.5 & 68.7 & 32.1 & 14.7 & 86.6 & 46.3 & 74.6 & 54.2 & 41.6 & 24.4 & 17.9 & 21.9 & 47.9 & 28.2 \\
    \multicolumn{1}{c|}{} & Basak (MICCAI'22) & 38.73 & 31.76 & 68.8 & 59.0 & 54.2 & 29.0 & 0.0 & 83.7 & 39.3 & 61.7 & 52.1 & 34.6 & 0.0 & 0.0 & 26.8 & 45.7 & 26.2 \\
    \multicolumn{1}{c|}{} & CLD (MICCAI'22) & 46.10 & 15.86 & 67.2 & 68.5 & 71.4 & 41.0 & 21.0 & 76.1 & 42.4 & 69.8 & 52.1 & 37.9 & 24.7 & 23.4 & 22.7 & 38.1 & 35.2 \\
    \multicolumn{1}{c|}{} & DHC (MICCAI'23)  & 49.53 & 13.89 & 68.1 & 69.6 & 71.1 & \textbf{42.3} & 37.0 & 76.8 & 43.8 & 70.8 & 57.4 & 43.2 & 27.0 & 28.7 & 29.1 & 41.4 & \textbf{36.7} \\
    \multicolumn{1}{c|}{} & A\&D (NeurIPS'23)  & 50.03 & 5.21 & 73.1 & 76.0 & 76.5 & 29.1 & 44.9 & 82.5 & 49.0 & 72.8 & 61.7 & 48.5 & 30.2 & 19.7 & 36.4 & 32.9 & 18.2 \\
    \multicolumn{1}{c|}{} & SKCDF (CVPR'25) & 53.81 & 5.97 & 77.1 & 77.9 & 71.2 & 34.1 & {50.4} & {88.6} & 51.6 & 80.9 & 58.9 & 48.8 & 33.0 & 30.2 & 32.2 & 45.9 & 26.4 \\
    \rowcolor{cyan!20} \multicolumn{1}{c|}{} & DINO-Mix (Ours) & \textbf{62.73} & \textbf{3.28} & \textbf{85.9}& \textbf{85.6}& \textbf{85.4}& 39.2&\textbf{59.6}& \textbf{90.1}& \textbf{62.5}& \textbf{86.9}& \textbf{74.2}& \textbf{60.2}& \textbf{44.0}& \textbf{38.9}& \textbf{44.1}& \textbf{50.3}& 34.1 \\
    
    \bottomrule
  \end{tabular}}
  \label{with sota on amos}
\end{table*}

\begin{figure*}[!t]
    \centering
    \includegraphics[width=0.95\linewidth]{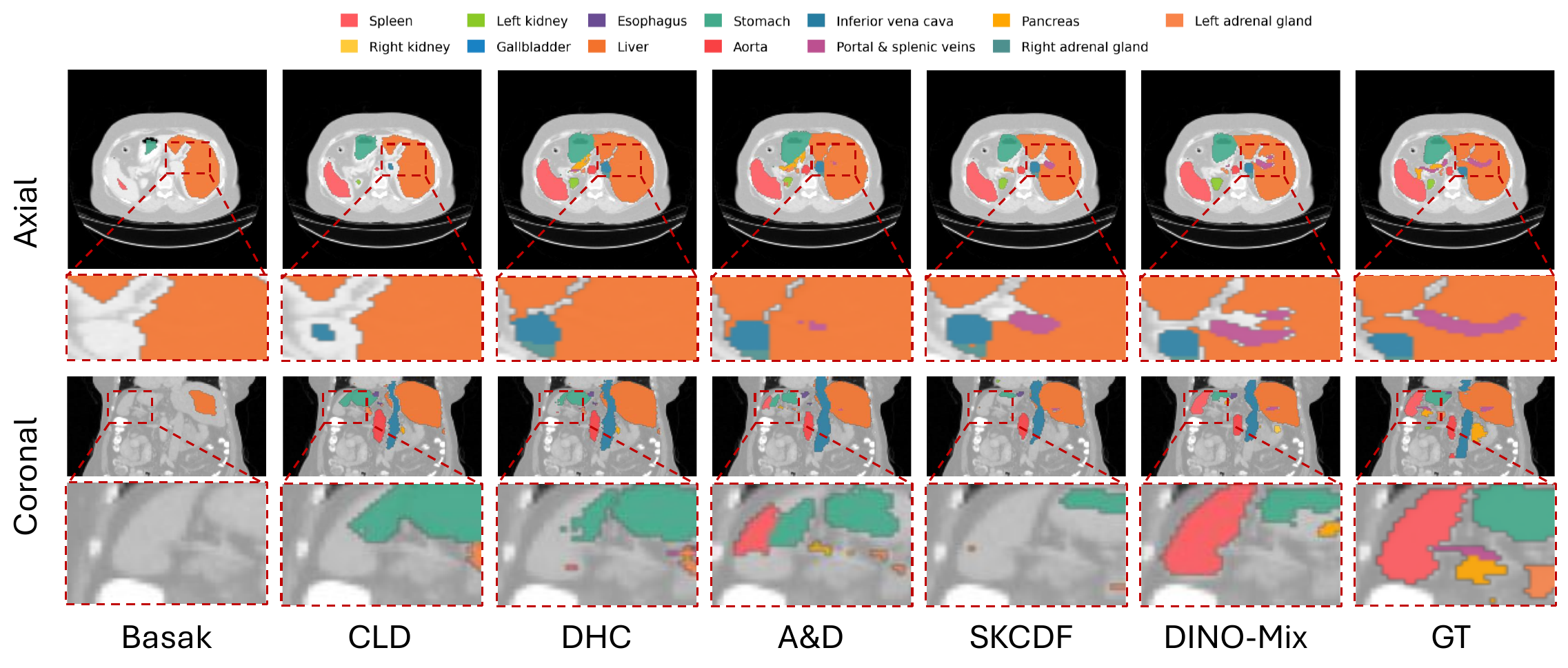}
    \caption{Qualitative comparison between our proposed DINO-Mix and the state-of-the-art semi-supervised class-imbalanced medical image segmentation methods on 20\% labeled Synapse dataset.}
    \label{fig:placeholder}
    \vspace{-12pt}
\end{figure*}

\subsection{Implementation Details}

We conduct our experiments using PyTorch 2.4.0, CUDA 12.4. The network is a simple VNet architecture \cite{vnet}, with parameters optimized with SGD with a momentum of 0.9 and an initial learning rate (lr) of 0.1 with momentum 0.9 and weight decay 3e-5. In the training stage, we randomly crop a volume of size 64 × 128 × 128. Random gamma augmentation is applied as the additional strong augmentation, and the ema weight is set to 0.99. The $\lambda_{unsup}$ is set to 0.5 during training. We train the networks for $E_{max}=$ 1500 epochs with a batch size of 4, consisting of 2 labeled and 2 unlabeled data. 
In the inference stage, only the VNet \cite{vnet} is utilized for prediction, thus \textit{no additional computation is required}. Final segmentation results are obtained using a sliding window strategy with a stride size of 32 × 32 × 16. For evaluation, we utilize two common metrics: the Dice Score (\%) and the Average Surface Distance (ASD) in voxels. The Dice Score quantifies the percentage of overlap between the predicted segmentation and the ground truth, while the ASD measures the average distance between their respective boundaries.

\subsection{Comparison with State-of-the-art Methods}

We rigorously compared our method against a suite of state-of-the-art semi-supervised segmentation techniques \cite{yu2019uamt, luo2021urpc, chen2021cps, wu2022ssnet, chen2022dst, wang2022depl}, including those specifically developed to address class-imbalance challenges \cite{guo2022adsh, wei2021crest, simis, basak2022addressing, lin2022cld, wang2023dhc, wang2023towards, zhang2025semantic}.

Our initial evaluation was performed on the Synapse dataset with a challenging scenario of only 20\% labeled data. As detailed in Table \ref{sota_IBSSL}, general semi-supervised methods have poor segmentation performance on small organs. Existing class-imbalance semi-supervised medical image segmentation methods yielded some improvements for minority classes but still struggle with the inherent class imbalance. Compared with them, our proposed approach demonstrates clear superiority. Specifically, DINO-Mix achieves a new state-of-the-art average Dice score of 66.45\% and an average ASD of 1.27, significantly outperforming the next-best method, SKCDF (64.27\% Dice, 1.45 ASD). This represents an absolute improvement of 2.18\% in Dice and a relative reduction of 12.4\% in ASD, demonstrating the enhanced accuracy and boundary precision of our model. Notably, DINO-Mix shows marked improvements in segmenting complex structures, boosting the Dice score for the Portal \& Splenic Veins (PSV) by 6.8\% (from 41.8\% to 48.6\%) and the Spleen (Sp) by 9.5\% (from 79.5\% to 89.0\%) compared to the previous state-of-the-art. Furthermore, it is observed that our method shows suprisingly robust performance across multiple runs, with merely 0.1 standard deviation on both Dice and ASD, which could suggest that our proposed FKD enables the model to learn robust representations.

\begin{table*}[t]
\caption{Ablation of the proposed components on the Synapse dataset with 20\% labeled data. Consist means the proposed consistency regularization based baseline framework. Results of 3-times repeated experiments are reported.}
  \footnotesize
  \setlength{\tabcolsep}{1.3mm}
\label{abl_components}
\resizebox*{\linewidth}{!}{
\begin{tabular}{cccc|cc|ccccccccccccc}
\toprule
\multirow{2}{*}{Consist} &  \multirow{2}{*}{FKD} & \multirow{2}{*}{PIC} & \multirow{2}{*}{Aux} & {Avg.} & {Avg.} & \multicolumn{13}{c}{Dice of Each Class}                                        \\ 
\multicolumn{4}{c|}{}                          &   Dice                         &   ASD                        & Sp   & RK   & LK   & Ga   & Es   & Li   & St   & Ao   & IVC  & PSV  & PA   & RAG  & LAG  \\\midrule
\checkmark &  & & & {{48.14±1.2}}	&{{7.07±3.2}}	&76.1& 64.8& 60.3& 50.2& 0.0& 93.3& 45.0& 82.9& 76.5& 30.0& 39.7& 0.0& 7.1\\
\checkmark & \checkmark &&  & {{54.72±0.1}}	&{{6.12±0.2}}	& 79.2& 71.4& 54.5& 51.1& 53.3&92.5& 45.6& 82.7&75.8& 36.1& 39.1& 0.0& 30.2 \\
\checkmark & & \checkmark & & {{64.97±3.4}}	&{{1.39±0.2}}	&87.2&66.7& 70.5& 63.6& 63.3& 94.1& 60.1& 84.9& 77.1& 50.0& 45.3& 46.1& 35.6 \\
\checkmark & \checkmark & \checkmark & & {{66.31±0.2}}	&{{1.28±0.2}}	&90.2& 73.4& 73.5& 67.0& 64.4& 94.5& 62.1& 86.4& 79.8& 48.2& 45.9& 44.6& 31.9 \\
\rowcolor{cyan!20} \checkmark & \checkmark & \checkmark & \checkmark& \textbf{{66.45±0.1}}	&\textbf{{1.27±0.1}}	&{89.0}& {73.3}& {71.8}& {68.7}& {64.0}& {94.4}& 59.4& {86.2}& {79.0}& {48.6}& 45.6& {47.3}& 36.6 \\
\bottomrule

\end{tabular}}
\end{table*}

\begin{table}[t] 
    \centering 
    \caption{Performance comparison with varied scaling parameter $\eta$ in PIC on 20\% labeled Synapse dataset.} 
    \label{tab:performance_metrics} 
    \begin{tabular}{l | c c} 
        \toprule 
        $\eta$ & Avg. Dice & Avg. ASD \\
        \midrule 
        0 & $64.87 \pm 0.9$ & $ 3.02 \pm 0.5$ \\
        1/2 & $65.96 \pm 1.1$ & $1.34 \pm 0.2$ \\
        \rowcolor{cyan!20} 2/3 & $66.45 \pm 0.1$ & $1.27 \pm 0.1$ \\
        $\infty$ & $65.75 \pm 0.2$ & $1.57 \pm 0.2$ \\
        \bottomrule 
    \end{tabular}
    \vspace{-8pt}
\end{table}

To further affirm the robustness and generalizability of our framework, we extended our evaluation to the more complex AMOS dataset, using an even more restricted 5\% labeled data setting. The results, presented in Table \ref{with sota on amos}, reinforce the consistent advantages of our method. DINO-Mix once again sets a new performance benchmark with a leading average Dice score of 62.73\% and an ASD of 3.28. This constitutes a substantial improvement of 8.92\% in absolute Dice score over the strongest prior method (53.81\%). The performance gains are particularly pronounced for challenging organs that suffer from class imbalance, such as the Pancreas (Pa), which saw its Dice score increase by 11.4\% (from 48.8\% to 60.2\%), and the Stomach (St), with a 10.9\% increase (from 51.6\% to 62.5\%). These consistent and significant gains across two distinct datasets and under different limited-label scenarios highlight the effectiveness and reliability of our proposed DINO-Mix framework. We present a qualitative comparison of various methods in Fig. \ref{fig:placeholder}. Existing methods frequently exhibit tendencies toward producing false segmentation regions or ignoring smaller anatomical structures. Our proposed approach consistently yields more precise and anatomically faithful delineations for these challenging minority organs.

\subsection{Ablation Studies}

\subsubsection{Ablation on the components}

To validate the contribution of each component, we performed an ablation study on the Synapse dataset in Table~\ref{abl_components}. The consistency regularization baseline achieves a 48.14\% mean Dice and suffers from the segmentation on several minority classes, which suggests the issue of confirmation bias. Integrating FKD significantly improves performance to 54.72\% Dice by introducing an unbiased external teacher, which proves crucial for external knowledge in learning underrepresented classes. Separately, adding PIC yields a leap to 64.97\% Dice by enforcing a dynamic curriculum focused on minority classes. The combination of FKD and PIC further boost performance to 66.31\%, demonstrating their synergistic effect, where the PIC-enhanced student better absorbs the knowledge from the FKD teacher. Finally, including an auxiliary head \cite{lee2021abc,zhang2025semantic} brings a peak performance of {66.45\%} mean Dice, confirming that each component is a valuable contributor to DINO-Mix framework.

\subsubsection{Ablation on the scaling parameter $\eta$}
The scaling parameter $\eta$ dictates the speed of the transition from a class-balanced ($P_{bal}$) to a uniform ($P_{uni}$) sampling distribution. In Table \ref{tab:performance_metrics}, our ablation study on the Synapse dataset reveals the importance of a well-paced transition. The extreme cases, employing a purely uniform ($\eta = 0$) or a purely class-balanced ($\eta = \infty$) strategy throughout training, both yield suboptimal results. The purely uniform approach performed the worst, suggesting the necessity of an initial phase of balanced sampling. Conversely, relying solely on class-balancing hindered the model's ability to adapt to the natural data distribution. The optimal performance in terms of both Dice and ASD was achieved with $\eta = 2/3$. This result validates our PIC strategy, which allows the model to first learn robust features for all classes, including rare ones, before gradually shifting to an uniform distribution. This moderately paced transition strikes the most effective balance.

\begin{table}[t] 
    \centering 
    \setlength{\tabcolsep}{1.25mm}
    \caption{Performance comparison with different choices of foundational model on 20\% labeled Synapse dataset.} 
    \label{tab:performance_metrics_found} 
    \begin{tabular}{l | c c} 
        \toprule 
        Model & Avg. Dice & Avg. ASD \\
        \midrule 
        DINOv2 \cite{oquab2023dinov2} & $65.86 \pm 0.3$ & $ 1.41 \pm 0.2$ \\
        \rowcolor{cyan!20} DINOv3 \cite{simeoni2025dinov3} & $66.45 \pm 0.1$ & $1.27 \pm 0.1$ \\
        CLIP \cite{clip} & $64.26 \pm 1.0$ & $1.49 \pm 0.7$ \\
        SigLIP2 \cite{tschannen2025siglip} & $65.72 \pm 0.1$ & $1.32 \pm 0.4$ \\
        \bottomrule 
    \end{tabular}
\end{table}

\subsubsection{Ablation on the choice of foundational model}
We ablated four prominent foundational models and summarize the results in Table \ref{tab:performance_metrics_found}. DINOv3 yields the best performance, achieving a superior Dice score of $66.45\%$ and the lowest Average ASD of $1.27$. While DINOv2 and SigLIP2 also proved to be effective, they did not reach the performance level of DINOv3. The vision-language model CLIP resulted in the lowest performance ($64.26\%$ Dice) and showed considerable instability. We hypothesize that the purely visual, self-supervised pre-training objective of the DINO models enables them to learn more generalizable and robust feature representations of fundamental concepts like shape and texture, which are critical for providing an unbiased supervisory signal in medical imaging. In contrast, the vision language models may be more aligned with textual semantics, making them less suitable for this dense prediction task. Therefore, we selected DINOv3 as the frozen teacher encoder for all experiments.

\section{Conclusion}

In this work, we addressed the critical failure of conventional "inward-looking" semi-supervised frameworks, which amplify confirmation bias on imbalanced medical datasets. We proposed DINO-Mix, a novel "outward-looking" paradigm that dismantles this vicious cycle on two synergistic levels. Through Foundational Knowledge Distillation (FKD), we leveraged an external, unbiased DINOv3 teacher to provide a stable semantic anchor for minority classes, defining what to learn. Simultaneously, our Progressive Imbalance-aware CutMix (PIC) curriculum cultivated a more robust student model, mastering how to learn by first confronting and then generalizing from class imbalances. Our state-of-the-art results on the Synapse and AMOS benchmarks validate this two-pronged strategy, demonstrating the effectiveness of looking outward to harness objective, foundational knowledge for reliable semi-supervised learning.

\bibliography{aaai2026}


\end{document}